\def\astroph{1}

\ifnum\astroph=0
\documentclass[manuscript]{aastex}          % Single column, double spaced
\else
\documentclass{emulateapj}
\usepackage{apjfonts}
\fi

\usepackage{epsfig,amsmath,natbib}

\newcommand{\fig}[1]{Fig.\ \ref{#1}}
\newcommand{\Fig}[1]{Figure \ref{#1}}

\begin{document}

\shorttitle{Collisionless Shocks}
\shortauthors{J. Trier Frederiksen, C. B. Hededal, T. Haugb\o lle, \AA . Nordlund}

\title{Magnetic Field Generation in Collisionless Shocks; Pattern Growth and Transport}
\author{J. Trier Frederiksen, C. B. Hededal, T. Haugb\o lle, \AA . Nordlund}
\affil{Niels Bohr Institute, Dept. of Astrophysics, Juliane Maries Vej 30, 2100 K\o benhavn \O , Denmark}
\email{trier@astro.ku.dk}

\begin{abstract}

We present results from three-dimensional particle simulations of collisionless 
shock formation, with relativistic counter-streaming ion-electron plasmas.  
Particles are followed over many skin depths downstream of the shock.  Open 
boundaries allow the experiments to be continued for several particle crossing 
times.  The experiments confirm the generation of strong magnetic and electric 
fields by a Weibel-like kinetic streaming instability, and demonstrate that the
electromagnetic fields propagate far downstream of the shock.  The magnetic 
fields are predominantly transversal, and are associated with merging ion current 
channels.  The total magnetic energy grows as the ion channels merge, and as the
magnetic field patterns propagate down stream.  The electron populations are 
quickly thermalized, while the ion populations retain distinct bulk speeds in
shielded ion channels and thermalize much more slowly.  The results help reveal 
processes of importance in collisionless shocks, and may help to explain the origin 
of the magnetic fields responsible for afterglow synchrotron/jitter radiation from 
Gamma-Ray Bursts.

\end{abstract}

\subjectheadings{acceleration of particles,
 gamma rays: bursts, instabilities, magnetic fields,
 plasmas}

\section{Introduction}

The existence of a strong magnetic field in the shocked external
medium is required in order to explain the observed radiation in
Gamma-Ray Burst afterglows as synchrotron radiation 
\citep[e.g.][]{bib:Panaitescu+Kumar}.  Nearly collisionless shocks,
with synchrotron-type radiation present, are also common in many other
astrophysical contexts, such as in super-nova shocks, and in jets
from active galactic nuclei.  At least in the context of Gamma-Ray
Burst afterglows the observed synchrotron radiation requires the
presence of a stronger magnetic field than can easily be explained by just
compression of a magnetic field already present in the external medium.

\citet{bib:MedvedevLoeb} showed through a linear kinetic treatment how a
two-stream magnetic instability -- a generalization of the Weibel
instability \citep{bib:Weibel,bib:YoonDavidson} -- can generate a
strong magnetic field  ($\epsilon_B$, defined as the ratio of magnetic energy to 
total kinetic energy, is $10^{-5}$-$10^{-1}$ of equipartition value) 
in collisionless shock fronts
\citep[see also discussion in][]{bib:RossiRees}. We
note in passing that this instability is well-known in other plasma
physics disciplines, e.g. laser-plasma interactions
\citep{bib:YangGallantAronsLangdon,bib:califano1},
and has been applied in the context of pulsar winds
by \citet{bib:Kazimura}.

Using three-dimensional particle-in-cell simulations to study
relativistic collisionless shocks (where an external plasma impacts the
shock region with a bulk Lorentz factor $\Gamma = 5-10$),
\cite{bib:astro-ph/0303360}, \cite{bib:Nishikawa}, and \cite{bib:Silva}
investigated the generation  of magnetic fields by the two-stream
instability.
In these first studies the
growth of the transverse scales of the magnetic field was limited by the
dimensions of the computational domains.  The durations of the 
\cite{bib:Nishikawa} experiments were less than particle travel 
times through the experiments, while \cite{bib:Silva} used periodic 
boundary conditions in the direction of streaming.
Further, \cite{bib:astro-ph/0303360} and \cite{bib:Nishikawa} used electron-ion ($e^-p$) 
plasmas, while experiments reported upon by \cite{bib:Silva} were done with $e^-e^+$
pair plasmas.

Here, we report on 3D particle-in-cell simulations of relativistically counter-streaming $e^-p$
plasmas. Open boundaries are used in the streaming direction, and experiment 
durations are several particle crossing times. Our results can help reveal the most important
processes in collisionless shocks, and help explain the observed afterglow
synchrotron radiation from Gamma-Ray Bursts. We focus on the earliest development in
shock formation and field generation. Late stages in shock formation will be
addressed in successive work. 

\section{Simulations}
Experiments were performed using a self-consistent 3D3V electromagnetic
particle-in-cell code originally developed for simulating reconnection 
topologies \citep{bib:HesseKuzenova}, 
redeveloped by the present authors to obey special relativity
and to be second order accurate in both space and time.

The code solves Maxwell's equations for the electromagnetic
field with continuous sources, with fields and field source terms defined on a staggered
3D Yee-lattice \citep{bib:Yee}. The sources in Maxwell's equations
are formed by weighted averaging of particle data to the field grid,
using quadratic spline interpolation. Particle velocities and positions are
defined in continuous (${\bf{r}},\gamma{\bf{v}}$)-space, and particles obey the relativistic 
equations of motion.

The grid size used in the main experiment was $(x,y,z)=200\times200\times800$,
with 25 particles per cell, for a total of $8\times10^8$ particles,
with ion to electron mass ratio $m_{i}/m_{e} = 16$.
To adequately resolve a significant number of electron and ion
skin-depths ($\delta_e$ and $\delta_i$), the box size was chosen such that
$L_{x,y} = 10\delta_i \sim 40\delta_e$ and $L_z \sim 40 \delta_i
\sim 160\delta_e$. Varying aspect and mass ratios were used in complementary experiments.

\begin{figure*}[!th]
\begin{center}
\epsfig{figure=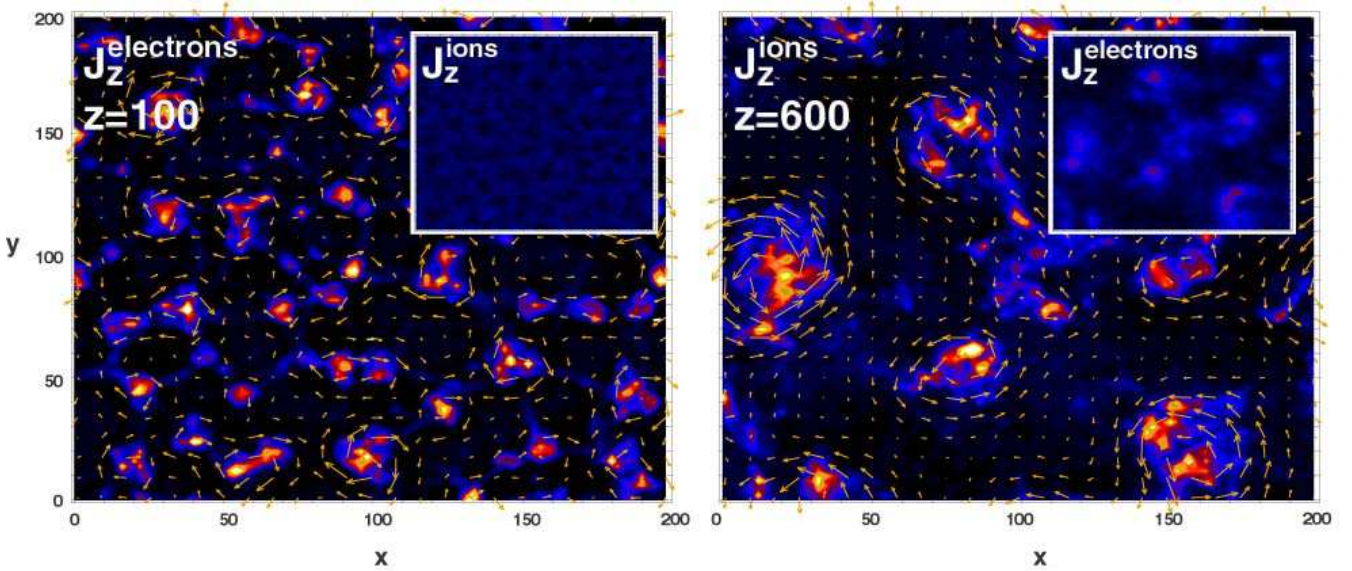,width=18cm}
\caption{
The left hand side panel shows the longitudinal electron current
density through a transverse cut at $z=100$, with a small inset showing
the ion current in the same plane.  The right hand side panel shows
the ion current at $z=600=30\delta_i$, with the small inset now instead showing
the electron current. The arrows represent the transverse magnetic field. Both panels are from time $t = 1200$.}
\label{fig:Slice}
\end{center}
\end{figure*}

Two counter-streaming -- initially
quasi-neutral and cold -- plasma populations are simulated. At the two-stream interface (smoothed around $z=80$)
a plasma ($z<80$) streaming in the positive z-direction, with a
bulk Lorentz factor $\Gamma=3$, hits another plasma ($z\ge80$) at rest in
our reference frame. The latter plasma is denser than the former by a factor of 3. 
Experiments have been run with both initially sharp and initially smooth 
transitions, with essentially the same results.  
The long simulation time gradually allows the shock to converge towards 
self-consistent jump conditions.
Periodic boundaries are imposed in
the $x$-- and $y$--directions, while the boundaries at $z=0$ and $z=800$ are open,
with layers absorbing transverse electromagnetic waves.  Inflow
conditions at $z=0$ are fixed, with incoming particles supplied at a
constant rate and with uniform speed.  At $z=800$ there is free outflow of particles.
The maximum experiment duration is 480 $\omega_{pe}^{-1}$ (where $\omega_{pe}$ is the electron plasma frequency), 
sufficient for propagating $\Gamma \approx 3$ particles 2.8 times through the box.

\section{Results and Discussions}

The extended size and duration of these experiments make it possible
to follow the two-stream instability through several stages of development;
first exponential growth, then non-linear saturation, followed by pattern growth 
and downstream advection.  We identify the mechanisms 
responsible for these stages below.

\subsection{Magnetic Field Generation, Pattern Growth \\ and Field Transport} \label{field_generation}
Encountering the shock front the
incoming electrons are rapidly (being lighter than the ions) deflected by
field fluctuations growing due to the two-stream instability \citep{bib:MedvedevLoeb}. 
The initial perturbations grow
non-linear as the deflected electrons collect into first caustic
surfaces and then current channels (\fig{fig:Slice}). Both streaming and 
rest frame electrons are deflected, by arguments of symmetry.
 
In accordance with Ampere's law the current channels
are surrounded by approximately cylindrical magnetic fields
(illustrated by arrows in \fig{fig:Slice}), causing
mutual attraction between the current channels.  The current
channels thus merge in a race where larger electron
channels consume smaller, neighboring channels.
In this manner, the transverse magnetic field
grows in strength and scale downstream. This continues until
the fields grow strong enough to deflect the
much heavier ions into the magnetic voids between
the electron channels. The ion channels are then subjected to
the same growth mechanism as the electrons. When ion channels
grow sufficiently powerful, they begin to experience Debye shielding by the
electrons, which by then have been significantly heated by scattering 
on the increasing electromagnetic field structures. The two electron 
populations, initially separated in $\gamma{\bf{v}}$-space, merge 
to a single population in approximately $20\delta_e$ ($z=80$--$200$) 
as seen in \fig{fig:acc}. The same trend is seen for the ions -- albeit 
at a rate slower in proportion to $m_i/m_e$.

\begin{figure}[!t]
\begin{center}
\epsfig{figure=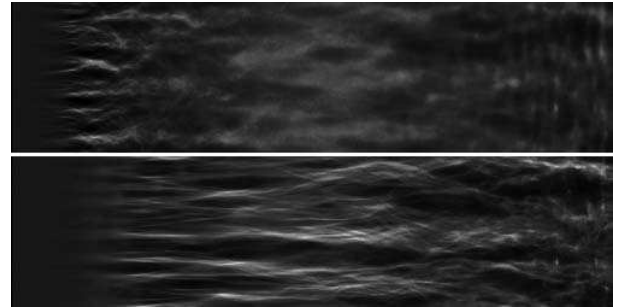,width=8cm}
\caption{
Electron (top) and ion (bottom) currents, averaged over the $x$-direction, at time
$t=1200$.
}
\label{fig:jiz}
\end{center}
\end{figure}
\begin{figure}[!ht]
\begin{center}
\epsfig{figure=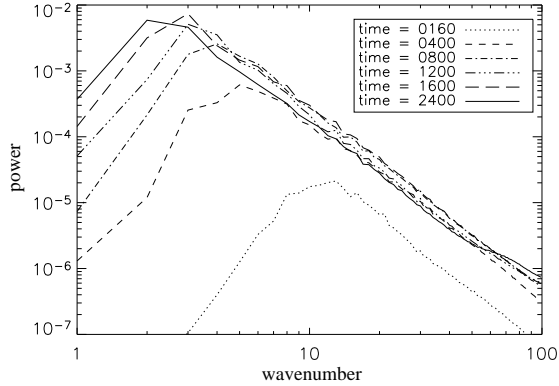,width=8.0cm}
\caption{Power spectrum of ${\mathbf B}_{\perp}$ for $z = 250$ at different times.}
\label{fig:power}
\end{center}
\end{figure}

The Debye shielding quenches the electron channels, while
at the same time supporting the ion-channels; the large
random velocities of the electron population allow the concentrated 
ion channels to keep sustaining strong magnetic fields.
\fig{fig:Slice}, shows the highly concentrated ion currents, the more diffuse
-- and shielding -- electron currents, and the resulting magnetic field.
The electron and ion channels are further illustrated in \fig{fig:jiz}.  
Note the limited $z$-extent of the electron
current channels, while the ion current channels extend throughout
the length of the box, merging to form larger scales downstream.
Because of the longitudinal current channels the magnetic field is 
predominantly transversal; we find $|B_z|/|B_{tot}| \sim 10^{-1} - 10^{-2}$.

\begin{figure}[!t]
\begin{center}
\epsfig{figure=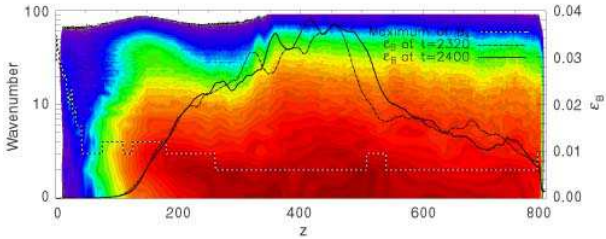,width=8.0cm}
\caption{Relative electromagnetic energy density $\epsilon_{B}$.
The contour color plot shows the power in the transverse magnetic
field through the box distributed on spatial Fourier modes at $t=2400$,
with the dotted line marking the wavenumber with maximum power.
Superposed is the spatial distribution of $\epsilon_{B}$, averaged across the beam,
at $t=2320$ (dashed-dotted) and $t=2400$ (full drawn), highlighting how EM-fields 
are advected down through the box.
}
\label{fig:epsb}
\end{center}
\end{figure}

\Fig{fig:power} shows the temporal development of the 
transverse magnetic field scales around $z=250$.
The power spectra follow power-laws, 
with the largest scales growing with time.
The dominant scales at these $z$ are of the order $\delta_i$
at early times. Later they become comparable to $L_{x,y}$. \Fig{fig:epsb} 
captures this scaling behavior as a function of depth for $t=2400$.

\begin{figure}[!t]
\begin{center}
\epsfig{figure=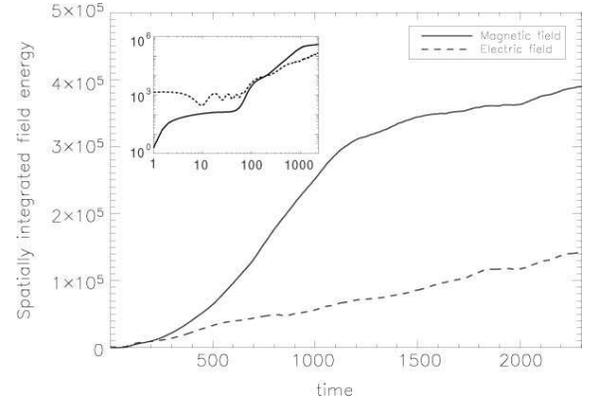,width=7.5cm}
\caption{
Total magnetic (full drawn) and electric (dashed) energy
in the box as a function of time. The inset shows a log-log 
plot of the same data.
}
\label{fig:B_energy}
\end{center}
\end{figure}

The time evolutions of the electric and magnetic field energies are shown in 
\fig{fig:B_energy}. Seeded by fluctuations in the fields, mass 
and charge density,  the two-stream instability initially grows super-linearly 
($t=80-100$), reflecting approximate exponential growth in a small sub-volume. Subsequently the 
total magnetic energy grows more linearly, reflecting essentially the 
increasing volume filling factor as the non-linearly saturated magnetic 
field structures are advected downstream.

At $t\approx 1100$ the slope drops off, due to advection of the generated
fields out of  the box.  The continued slow growth, for $t > 1100$, reflects
the increase of the pattern size with time (cf.\ \fig{fig:power}).  
A larger pattern size
corresponds to, on the average, a larger mean magnetic energy, since the
total electric current is split up into fewer but stronger ion current 
channels.  
The magnetic energy scales with
the square of the electric current, which in turn grows in inverse proportion
to the number of current channels.  The net
effect is that the mean magnetic energy increases accordingly.

The magnetic energy density keeps growing throughout our experiment,
even though the duration of the experiment (480 $\omega_{pe}^{-1}$) significantly 
exceeds the particle crossing time, and also exceeds the advection time of the 
magnetic field structures through the box.  This is in contrast to the results 
reported by \citet{bib:Silva}, 
where the magnetic energy density drops back after about 10-30 $\omega_{pe}^{-1}$.
It is indeed obvious from the preceding discussion that the ion-electron 
asymmetry is essential for the survival of the current channels.

From the requirement that the total plasma momentum should be conserved,
the (electro)magnetic field produced by the two-stream instability
acquires part of the z-momentum lost by the two-stream population
in the shock; this opens the possibility that magnetic field structures
created in the shock migrate downstream of the shock and thus
carry away some of the momentum impinging on the shock.

Our experiments show that this does indeed happen; 
the continuous injection of momentum transports the generated
field structures downstream at an accelerated advection speed.
The dragging of field structures through the dense plasma acts as to transfer momentum between 
the in-streaming and the shocked plasmas.

\subsection{Thermalization and Plasma Heating}
\begin{figure}[!t]
\begin{center}
\epsfig{figure=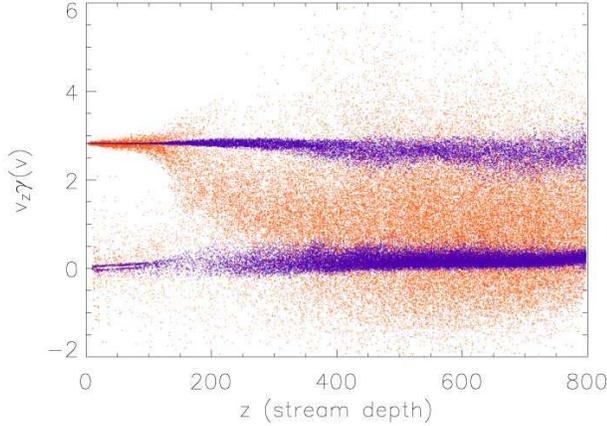,width=8cm}
\caption{Thermalization and longitudinal acceleration,
illustrated by scatter plots of the electron (orange) and ion (blue) 
populations. 
Note the back-scattered electron population ($v_z\gamma(v) < 0$).
}
\label{fig:acc}
\end{center}
\end{figure}

At late times the entering electrons are effectively
scattered and thermalized: The magnetic field isotropizes the velocity distribution 
whereas the electric field generated by the $e^{-}$--$p$ charge separation acts to thermalize the populations.
\Fig{fig:acc} shows that this happens over the $\sim$ 20 electron skin
depths from around $z=80$ -- $200$. 
The ions are expected to also thermalize, given sufficient space and time. This
fact leaves the massive ion bulk momentum constituting a vast energy reservoir for
further electron heating and acceleration. Also seen in \fig{fig:acc}, the ions
beams stay clearly separated in phase space,  and are only slowly broadened (and
heated).

We do not see indications of a super-thermal tail in the heated electron
distributions, and there is thus no sign of second order Fermi-acceleration in 
the experiment presented in this Letter.
\cite{bib:Nishikawa} and \cite{bib:Silva} reported acceleration of particles 
in experiments similar to the current experiment, except for more limited
sizes and durations, and the use of an $e^-e^+$ plasma \citep{bib:Silva}.
On closer examination of the published results it appears that there is no
actual disagreement regarding the absence of accelerated particles. Whence,
\cite{bib:Nishikawa} refer to transversal velocities of the order of $0.2
c$ (their Fig.\ 3b), at a time where our experiment shows similar
transversal velocities (cf. \fig{fig:acc}) 
that later develop a purely thermal spectrum. \cite{bib:Silva} refer to 
transversal velocity amplitudes up to about $0.8 c$ (their Fig.\ 4), or $v\gamma\sim 2$,
with a shape of the distribution function that appears to be compatible with thermal.
In comparison, the electron distribution illustrated by the scatter plot in \fig{fig:acc} 
covers a similar interval of $v\gamma$, with distribution functions that are 
close to (Lorentz-boosted) relativistic Maxwellians.  Thus, there 
is so far no compelling evidence for non-thermal particle acceleration in experiments 
with no imposed external magnetic field. Thermalization is a more likely cause 
of the increases in transversal velocities.

\cite{bib:astro-ph/0303360} reported evidence for particle acceleration,
with electron gammas up to $\sim100$, in experiments 
with an external magnetic field present in the up-stream plasma.  This is indeed 
a more promising scenario for particle acceleration experiments 
\citep[although in the experiments by][results with an external magnetic field
were similar to those without]{bib:Nishikawa}.  \Fig{fig:acc}
shows the presence of a population of back-scattered electrons ($v_z\gamma < 0$).  In the presence of
an external magnetic field in the in-streaming plasma, this possibly facilitates Fermi acceleration in the shock.

\section{Conclusions}

The experiment reported upon here illustrates a number of fundamental properties
of relativistic, collisionless shocks:

1.
Even in the absence of a magnetic field in the up-stream plasma,
a small scale, fluctuating, and predominantly transversal magnetic field
is unavoidably generated by a two-stream instability reminiscent of the
Weibel-instability. In the current experiment the magnetic energy density 
reaches a few percent of the energy density of the in-coming beam.

2.
In the case of an $e^-p$ plasma the electrons are rapidly thermalized, while
the ions form current channels that are the sources of deeply
penetrating magnetic field structures.  The channels merge in the downstream 
direction, with a corresponding increase of the average
magnetic energy with shock depth.  This is expected
to continue as long as a surplus of bulk relative momentum remains in the 
counter-streaming plasmas.

3.
The generated magnetic field patterns are advected downstream at speeds 
intermediate of the streaming and restframe plasmas.
The electromagnetic field structures
thus provide scattering centers that interact with both the fast, in-coming
plasma, and with the plasma that is initially at rest.  As a result the 
electron populations of both components quickly thermalize and form
a single, Lorentz-boosted thermal electron population.  The two ion populations 
merge much more slowly, with only gradually increasing ion temperatures.

4. The observed strong turbulence in the field structures at the shocked streaming interface 
provides a promising environment for particle acceleration.

We emphasize that quantification of the interdependence and development of
$\epsilon_U$ and $\epsilon_B$ is accessible by means of such experiments as reported
upon here. 

Rather than devising abstract scalar parameters $\epsilon_B$
and $\epsilon_U$, that may be expected to depend on shock depth, media densities
etc., a better approach is to compute synthetic radiation spectra directly from 
the models, and  then apply scaling laws to predict what would be observed from 
corresponding, real supernova remnants and Gamma-Ray Burst afterglow shocks.
\vspace{6pt}

We are grateful to Dr. Michael Hesse / GSFC for generously providing the
original particle-in-cell code and for 
helpful discussions on the implementation and on numerical issues in 
particle simulations. Computer time was provided by the Danish Center for Scientific Computing.

\end{document}